\documentclass[12pt]{article}
 \usepackage{epsfig}
 \def\be{\begin{equation}}
 \def\ee{\end{equation}}
 \def\bea{\begin{eqnarray}}
 \def\eea{\end{eqnarray}}
 \usepackage{graphicx}

 \catcode`\@=11
 \def\lsim{\mathrel{\mathpalette\@versim<}}
 \def\gsim{\mathrel{\mathpalette\@versim>}}
 \def\@versim#1#2{\vcenter{\offinterlineskip
 \ialign{$\m@th#1\hfil##\hfil$\crcr#2\crcr\sim\crcr } }}
 \catcode`\@=12

 \catcode`@=12
\begin{document}
 \thispagestyle{empty}
 \begin{flushright}
 UCRHEP-T568\\
 June 2016\
 \end{flushright}
 \vspace{0.6in}
 \begin{center}
 {\LARGE \bf Progressive Gauge U(1) Family Symmetry\\
 for Quarks and Leptons\\}
 \vspace{1.2in}
 {\bf Ernest Ma\\}
 \vspace{0.2in}
 {\sl Physics \& Astronomy Department and Graduate Division,\\ 
 University of California, Riverside, California 92521, USA\\}
 \end{center}
 \vspace{1.2in}

\begin{abstract}\
The pattern of quark and lepton mass matrices is unexplained in the 
standard model of particle interactions.  I propose the novel idea of a 
progressive gauge $U(1)$ symmetry where it is a reflection of the 
regressive electroweak symmetry breaking pattern, caused by 
an extended Higgs scalar sector.  Phenomenological implications of 
this new hypothesis are discussed.
\end{abstract}

 \newpage
 \baselineskip 24pt

The standard model (SM) of particles interaction is based on the gauge 
symmetry $SU(3)_C \times SU(2)_L \times U(1)_Y$, with its associated 
vector gauge bosons, i.e. eight gluons, the weak $W^\pm$ and $Z^0$ bosons, 
and the photon.  It consists of three families of quarks and leptons in 
left-handed doublets and right-handed singlets.  It also has the 
all-important Higgs scalar doublet which provides mass directly 
to all particles with the possible exception of only the neutrinos. 
The resulting quark and lepton masses and their mixing patterns are 
unexplained in the SM.  They are merely tunable parameters.  To gain 
an understanding of these patterns, I propose that there is a family 
gauge $U(1)$ symmetry, which requires an extended scalar sector, which 
breaks this $U(1)$ as well as $SU(2)_L \times U(1)_Y$ in a regressive 
manner~\cite{m01} so that the observed patterns of quark and lepton 
masses and mixing are qualitatively explained.  

The fermion content of the SM is extended to include three singlet 
right-handed neutrinos $\nu_R$.  The new family gauge $U(1)_F$ symmetry 
is assumed coupled only to right-handed fermions, as shown in Table 1.
\begin{table}[htb]
\caption{Fermion assignments under $U(1)_F$.}
\begin{center}
\begin{tabular}{|c|c|c|c|c|}
\hline
Particle & $SU(3)_C$ & $SU(2)_L$ & $U(1)_Y$ & $U(1)_F$ \\
\hline
$Q_{Li} = (u,d)_{Li}$ & 3 & 2 & 1/6 & $(0,0,0)$ \\
$u_{Ri}$ & $3$ & 1 & $2/3$ & $(n_1,n_2,n_3)$ \\
$d_{Ri}$ & $3$ & 1 & $-1/3$ & $(-n_1,-n_2,-n_3)$ \\
\hline
$L_{Li} = (\nu,l)_{Li}$ & 1 & 2 & $-1/2$ & $(0,0,0)$ \\
$l_{Ri}$ & 1 & 1 & $-1$ & $(-n_1,-n_2,-n_3)$ \\
$\nu_{Ri}$ & 1 & 1 & 0 & $(n_1,n_2,n_3)$ \\
\hline
\end{tabular}
\end{center}
\end{table}
The $[SU(3)_C]^2 U(1)_F$ anomaly is cancelled between $u_R$ and $d_R$ for each 
family.  The $[SU(2)_L]^2 U(1)_F$ anomaly is zero because left-handed fermions 
do not couple to $U(1)_F$.  The $[U(1)_Y]^2 U(1)_F$ and $U(1)_Y [U(1)_F]^2$ 
anomalies are cancelled between $u_R$, $d_R$, and $l_R$ for each family.  
The $[U(1)_F]^3$ anomaly is canceled between $u_R$ and $d_R$, as well as 
$l_R$ and $\nu_R$ for each family.  This means that $U(1)_F$ is anomaly-free 
within each family (which is basically just $B-L-2Y$), but it may have an 
overall different coupling for each, as shown in Table 1.

To obtain quark and lepton masses, there should then be three Higgs doublets: 
\begin{equation}
\Phi_i = (\phi^+,\phi^0)_i \sim (1,2,1/2;n_i),
\end{equation}
coupling in turn to the three families. 
Let the scalar singlet $\sigma \sim [1,1,0;(n_2-n_3)/2]$ be added, then 
the scalar potential has the relevant terms
\begin{equation}
V = m_2^2 \Phi_2^\dagger \Phi_2 + \kappa_2 \sigma^2 \Phi_2^\dagger \Phi_3 
+ ... 
\end{equation}
If $m_2^2$ is positive and large, then the vacuum expectation value of 
$\phi_2^0$ is given by
\begin{equation}
v_2 \simeq {-\kappa_2 u^2 v_3 \over m_2^2},
\end{equation}
where $v_{2,3} = \langle \phi^0_{2,3} \rangle$, $u = \langle \sigma \rangle$, 
and $v_2$ may be small because $\kappa_2 \to 0$ enlarges the symmetry of $V$.  
This mechanism based on Ref.~\cite{m01} is easily generalized~\cite{glr09}. 
If $n_2-n_3=n_1-n_2$ as well, then the term $\kappa_1 \sigma^2 \Phi_1^\dagger 
\Phi_2$ also exists, so that
\begin{equation}
v_1 \simeq {-\kappa_1 u^2 v_2 \over m_1^2},
\end{equation}
which yields $v_3 >> v_2 >> v_1$ and explains the hierarchy of quark and 
lepton masses.  Thus the regressive pattern of electroweak symmetry breaking 
from $\Phi_{3,2,1}$ results in the progressive pattern of masses for the 
first, second, and third families.

As an explicit example, let $n_{1,2,3} = (2,1,0)$ with $\sigma \sim 
(1,1,0;1/2)$, then the most general scalar potential consisting of 
$\Phi_{1,2,3}$ and $\sigma$ is given by
\begin{eqnarray}
V &=& m_1^2 \Phi_1^\dagger \Phi_1 + m_2^2 \Phi_2^\dagger \Phi_2 + 
m_3^2 \Phi_3^\dagger \Phi_3 + m_4^2 \sigma^* \sigma \nonumber \\ 
&+& {1 \over 2} \lambda_1 (\Phi_1^\dagger \Phi_1)^2 + 
{1 \over 2} \lambda_2 (\Phi_2^\dagger \Phi_2)^2 + 
{1 \over 2} \lambda_3 (\Phi_3^\dagger \Phi_3)^2 + 
{1 \over 2} \lambda_4 (\sigma^* \sigma)^2 \nonumber \\ 
&+& \lambda_{12} (\Phi_1^\dagger \Phi_1)(\Phi_2^\dagger \Phi_2) + 
\lambda_{13} (\Phi_1^\dagger \Phi_1)(\Phi_3^\dagger \Phi_3) + 
\lambda_{23} (\Phi_2^\dagger \Phi_2)(\Phi_3^\dagger \Phi_3) \nonumber \\  
&+& \lambda'_{12} (\Phi_1^\dagger \Phi_2)(\Phi_2^\dagger \Phi_1) + 
\lambda'_{13} (\Phi_1^\dagger \Phi_3)(\Phi_3^\dagger \Phi_1) + 
\lambda'_{23} (\Phi_2^\dagger \Phi_3)(\Phi_3^\dagger \Phi_2) \nonumber \\  
&+& \lambda_{14} (\Phi_1^\dagger \Phi_1)(\sigma^* \sigma) + 
\lambda_{24} (\Phi_2^\dagger \Phi_2)(\sigma^* \sigma) + 
\lambda_{34} (\Phi_3^\dagger \Phi_3)(\sigma^* \sigma) \nonumber \\  
&+& [\lambda_{123} (\Phi_1^\dagger \Phi_2)(\Phi_3^\dagger \Phi_2) + 
\kappa_1 \sigma^2 \Phi_1^\dagger \Phi_2 + 
\kappa_2 \sigma^2 \Phi_2^\dagger \Phi_3 + H.c.] 
\end{eqnarray}
For large positive $m^2_{1,2}$ and negative $m^2_{3,4}$, the minimum of 
$V$ satisfies the conditions:
\begin{eqnarray}
&& m_4^2 + \lambda_4 |u|^2 + \lambda_{34} |v_3|^2 \simeq 0, \\ 
&& m_3^2 + \lambda_3 |v_3|^2 + \lambda_{34} |u|^2 \simeq 0, \\ 
&& v_2 \simeq {-\kappa_2 u^2 v_3 \over m_2^2 + (\lambda_{23} + \lambda'_{23}) 
|v_3|^2 + \lambda_{24} |u|^2}, \\
&& v_1 \simeq {-\kappa_1 u^2 v_2 \over m_1^2 + (\lambda_{13} + \lambda'_{13}) 
|v_3|^2 + \lambda_{14} |u|^2}.
\end{eqnarray}

This regressive pattern of electroweak symmetry breaking allows a 
qualitative understanding of why $m_{u,d} << m_{c,s} << m_{t,b}$.  The  
quark mass matrix linking $(\bar{d},\bar{s},\bar{b})_L$ to 
$(d,s,b)_R$ is of the form
\begin{equation}
{\cal M}_d = \pmatrix {U_{1d} & U_{1s} & U_{1b} \cr U_{2d} & U_{2s} & U_{2b} \cr 
U_{3d} & U_{3s} & U_{3b}} \pmatrix {m'_d & 0 & 0 \cr 0 & m'_s & 0 \cr 
0 & 0 & m'_b},
\end{equation}
where $m'_{d,s,b} \propto v_{1,2,3}$, and $\sum_i |U_{id}|^2 = \sum_i |U_{is}|^2 
= \sum_i |U_{ib}|^2 = 1$.  However, $\sum_i U^*_{id} U_{is}$, etc. are not 
necessarily zero, so $m'_{d,s,b}$ are not necessarily the mass eigenvalues. 
If $U_{1d} = U_{2s} = U_{3b} = 1$, ${\cal M}_d$ is diagonal.  Similarly 
if $U_{1u} = U_{2c} = U_{3t} = 1$, ${\cal M}_u$ is also diagonal.  This 
corresponds to the alignment limit where there is a separate global 
U(1) symmetry for each family.  Hence it is technically natural to 
expect the mixing between families to be small, i.e. $|U_{2d,3d}| << |U_{1d}|$, 
etc.  Note that this argument does not work in the SM, because there is 
no mechanism there to enforce the hierarchy of quark masses, so that an 
off-diagonal term in ${\cal M}_d$ for example may be bigger than $m_d$ 
itself.  Here the mass scale for each column of ${\cal M}_d$ is dictated by a 
specific hierarchical $v_i$.

Now ${\cal M}_d$ is diagonalized in general by
\begin{equation}
{\cal M}_d = U_{dL} \pmatrix{m_d & 0 & 0 \cr 0 & m_s & 0 \cr 0 & 0 & m_b} 
U^\dagger_{dR},
\end{equation}
where both $U_{dL}$ and $U_{dR}$ are unitary matrices and assumed here to be 
close to the identity matrix.  To minimize the appearance of flavor-changing 
neutral currents in the $U(1)_F$ sector, it will be assumed~\cite{km16} 
that $U_{dR} = U_{uR} = 1$.  As usual the charged-current 
mixing matrix in the electroweak sector is 
\begin{equation}
V_{CKM} = U_{uL}^\dagger U_{dL},
\end{equation}
and the neutral-current interaction through the $Z$ boson is diagonal 
and universal as in the SM.  Thus the gauge sector here is absent of 
tree-level flavor-changing neutral currents.
The mass-squared matrix spanning the $(Z,Z_F)$ gauge bosons is given by
\begin{equation}
{\cal M}^2_{Z,Z_F} = \pmatrix{ (1/2)g_Z^2(v_1^2+v_2^2+v_3^2) & 
g_Z g_F (2v_1^2 + v_2^2) \cr g_Z g_F (2v_1^2 + v_2^2) &  (1/2)g_F^2 u^2}.
\end{equation}
The mixing between $Z$ and $Z_F$ is of order $(2g_Z/g_F)(2v_1^2+v_2^2)/u^2$ 
which is very small, say at most $10^{-5}$ in this model, and may be safely 
neglected.  Using Table 1, the branching fraction of $Z_F$ to $e^-e^+ + \mu^-
\mu^+$ is about 1/8.  The $c_{u,d}$ coefficients used in the experimental 
search~\cite{atlas14_z,cms15_z} of $Z_F$ are then
\begin{equation}
c_u = c_d = 4 g_F^2 (1/8).
\end{equation}
For $g_F = 0.1$, a lower bound of about 3.0 TeV on $m_{Z_F}$ is obtained 
from the Large Hadron Collider (LHC) based on data from the 7 and 8 TeV 
runs.  In that case, the lower limit on $u$ is about 42.4 TeV.  If $Z_F$ 
is discovered, then it may be distinguished from other $Z'$ models by 
the ratio 
\begin{equation}
{\Gamma (Z_F \to e^- e^+) \over \Gamma(Z_F \to \mu^- \mu^+)} = 
{n_1^2 \over n_2^2} = 4.
\end{equation}

The particle spectrum of this model consists of the heavy vector gauge 
boson $Z_F$ as well as the heavy scalar $\Phi_{1,2}$ doublets and the 
heavy scalar $\sigma$ singlet.  The rest are 
just the SM particles, with the important difference that the SM Higgs 
boson is now replaced by a linear combination $h = \sum_i a_i h_i$, where 
$h_{1,2,3} = \sqrt{2} Re(\phi^0_{1,2,3})$.  [There may also be a $\sigma$ 
component which is assumed negligible in this study.  If it is included, 
then since $\sigma$ does not couple to the SM fermions, its effect is to 
reduce all $h$ couplings by an overall factor.]  This $h$  should of course 
be identified 
as the 125 GeV particle~\cite{atlas12,cms12} discovered at the LHC.  If 
$a_i = v_i (\sum_i v_i^2)^{-1/2}$, then $h$ is the SM Higgs boson.  If not, 
there could be significant deviations in the production and decay of $h$, as 
discussed below.  The couplings of $h_{1,2,3}$ to quarks and leptons are 
given by
\begin{eqnarray}
{\cal L}_Y &=& {1 \over \sqrt{2}} (\bar{u},\bar{c},\bar{t})_L U^\dagger_{uL} 
\pmatrix{m_u h_1/v_1 & 0 & 0 \cr 0 & m_c h_2/v_2 & 0 \cr 0 & 0 & 
m_t h_3/v_3}  \pmatrix{u \cr c \cr t}_R \nonumber \\ 
&+& {1 \over \sqrt{2}} (\bar{d},\bar{s},\bar{b})_L U^\dagger_{dL} 
\pmatrix{m_d h_1/v_1 & 0 & 0 \cr 0 & m_s h_2/v_2 & 0 \cr 0 & 0 & 
m_b h_3/v_3}  \pmatrix{d \cr s \cr b}_R \nonumber \\ 
&+& {1 \over \sqrt{2}} (\bar{e},\bar{\mu},\bar{\tau})_L U^\dagger_{lL} 
\pmatrix{m_e h_1/v_1 & 0 & 0 \cr 0 & m_\mu h_2/v_2 & 0 \cr 0 & 0 & 
m_\tau h_3/v_3} \pmatrix{e \cr \mu \cr \tau}_R + H.c. 
\end{eqnarray}
In the above, the left-handed fermions are not mass eigenstates.  They 
are rotated by the unitary $U_L$ matrices.  This is easily seen by 
replacing $h_{1,2,3}$ with $\sqrt{2} v_{1,2,3}$ in Eq.~(16), which reduces the 
coupling matrices to mass matrices.  The mismatch between the $up$ and 
$down$ sectors generates thus Eq.~(12).  Nevertheless, each $h_i$ couples 
diagonally to all fermions in their mass-eigenvalue bases, i.e. $h_1$ 
couples to $\bar{u}_L u_R$, $\bar{d}_L d_R$, $\bar{e}_L e_R$; $h_2$ 
couples to $\bar{c}_L c_R$, $\bar{s}_L s_R$, $\bar{\mu}_L \mu_R$; and $h_3$ 
couples to $\bar{t}_L t_R$, $\bar{b}_L b_R$, $\bar{\tau}_L \tau_R$.  
This is a remarkable result, because flavor-changing neutral-current 
couplings are supposed to be unavoidable in models with several Higgs 
doublets.  Its origin is the assumption $U_{uR} = U_{dR} = U_{lR} = 1$, 
which corresoponds to a symmetry limit.  An immediate 
prediction is that there is no $h \to \tau \mu$ coupling here in the 
mass-eigenvalue basis of charged leptons.  If the preliminary 
indication~\cite{cms15_tau} of a nonzero branching fraction for 
this process is confirmed, this assumption must be relaxed.

Let $a_i = x_i v_i (\sum_i v_i^2)^{-1/2}$ with $\sum_i a_i^2 = 1$.  Then for 
$x_i \neq 1$, there are possible observable deviations from the SM in 
Higgs interactions.  For example, for $x_3 \neq 1$, the production of 
$h$ through the $t$ and $b$ quark loops in gluon fusion is changed by 
the factor $x_3^2$.  This is probably a small effect, because $v_3$ 
dominates over $v_{2,1}$, so $v_3$ is still very close to the SM value of 
$v = \sqrt{v_3^2+v_2^2+v_1^2}$.  However, $h$ decay to the second and 
first families may be strongly affected.  For example, for $v=174$ GeV, 
let $v_1 = 0.5$ GeV, $v_2 = 10$ GeV, then $v_3 = 173.7$ GeV, and 
$v_3^2/v^2 = 0.9967$ and $v_2^2/v^2 = 0.0033$.  Now 
\begin{equation}
x_3^2 = 1.0033 (1 - 0.0033 x_2^2),
\end{equation}
where $v_1^2/v^2 = 8.3 \times 10^{-6}$ has been neglected.  The SM 
limit is $x_2=x_3=1$, but $x_2$ may easily be much larger, e.g. $x_2=2$ 
and $x_3 = 0.995$.  Whereas the effect of the small deviation of $x_3$ 
from unity is very hard to observe, the consequence of a large $x_2$ is 
potentially observable in $h \to \mu^- \mu^+$ which would be enhanced 
by a factor of $x_2^2$.  At present the LHC bounds~\cite{atlas14_mu,cms15_mu} 
are about 7 times the SM value at 95\% CL, hence $x_2 < 2.6$ is allowed.
If $h \to \mu^- \mu^+$ is indeed observed at a rate much above the SM 
prediction, then in this model, the same must be true for $h \to c \bar{c}$. 
The SM prediction for $h \to c \bar{c}$ is about 2.5\%, but it is obscured 
by a large background from the strong production of charm quarks.  If it is 
enhanced by $x_2^2$, it may then be marginally observable~\cite{dgps14}.

\noindent \underline {In summary, the $h$ of this model couples diagonally 
to all fermions as in the standard model,}\\
\underline {but differs from it by having an additional factor $x_i \neq 1$ 
for each family.} 

In the neutrino sector, the $3 \times 3$ Yukawa coupling matrix linking 
$\bar{\nu}_{iL}$ to $\nu_{jR}$, again with the assumption $U_{\nu R} = 1$, 
is given by
\begin{equation}
{\cal L}_\nu = {1 \over \sqrt{2}} (\bar{\nu}_e,\bar{\nu}_\mu,\bar{\nu}_\tau)_L 
U^\dagger_{\nu L} \pmatrix{m_{D1} h_1/v_1 & 0 & 0 \cr 0 & m_{D2} h_2/v_2 & 0 
\cr 0 & 0 & m_{D3} h_3/v_3} \pmatrix{\nu_e \cr \nu_\mu \cr \nu_\tau}_R + H.c. 
\end{equation}
Adding a scalar singlet $\sigma' \sim (1,1,0;3)$ to break $U(1)_F$, the 
$3 \times 3$ Majorana mass matrix spanning $(\nu_{e,\mu,\tau})_R$ is 
of the form
\begin{equation}
{\cal M}_\nu = \pmatrix{0 & M_3 & 0 \cr M_3 & 0 & 0 \cr 0 & 0 & M_0},
\end{equation}
where $M_0$ is an allowed mass term, and $M_3$ comes from 
$\langle \sigma' \rangle$.  Hence the mismatch between $U_{lL}$ and 
$U_{\nu L}$ generates the neutrino mixing matrix, whereas the neutrino mass 
eigenvalues are $\pm m_{D1} m_{D2}/M_3$ and $-m_{D3}^2/M_0$.  This 
approximates the realistic case of two almost degenerate neutrinos 
for solar oscillations and one other for atmospheric oscillations. 
The splitting of the two degenerate masses may be achieved with a 
slight relaxation of the $U_{\nu R} = 1$ assumption for example. 
The addition of $\sigma'$ means that the mass of $Z_F$ gets another 
contribution.  It is now given by
\begin{equation}
m^2_{Z_F} = {1 \over 2} g_F^2 (u^2 + 36 {u'}^2).
\end{equation}
This allows a smaller value for $u$, say 1 TeV, in Eqs.~(3) 
and (4).  Setting $m_{Z_F} > 3$ TeV for $g_F = 0.1$, 
a lower limit $u' > 7$ TeV is obtained.   To connect $\sigma'$ to $V$ 
of Eq.~(5), another scalar $\sigma'' \sim (1,1,0;-3/2)$ may be added to 
allow the terms $\sigma' (\sigma'')^2$ and $\sigma'' \sigma^3$.

In conclusion, a progressive gauge $U(1)_F$ family symmetry is proposed 
for quarks and leptons.  It is anomaly-free within each family, but it has 
a different overall coupling for each, as shown in Table 1.  Three scalar 
doublets $\Phi_{1,2,3}$ are required, coupling each to a different family. 
The regressive pattern of electroweak symmetry breaking, i.e. $v_3 >> v_2 
>> v_1$, as shown in Eqs.~(6) to (9), offers an understanding to the 
observed hierarchy of fermion masses, 
i.e. $m_u << m_c << m_t$, and $m_d << m_s << m_b$, and $m_e << m_\mu << m_\tau$. 
Since each family has its own mass scale, the limit of no mixing is 
technically natural because it corresponds to an extra global U(1) symmetry.  
This is a possible explanation of the observed small mixing in the quark 
sector.  In the lepton sector, because of the additional Majorana mass 
matrix of the right-handed neutrinos, this limit of no mixing is spoiled. 
Hence small mixing is not expected.  The two main predictions of this new 
proposal are:
\begin{itemize}
\item {There exists a $Z_F$ gauge boson which couples right-handedly to 
the three families of quarks and leptons with different overall couplings. 
In particular, the ratio $\Gamma (Z_F \to e^- e^+)/\Gamma (Z_F \to \mu^- \mu^+)$ 
is $n_1^2/n_2^2$ which is in general not equal to one.  In the example studied 
in this paper, it is 4.  For $g_F = 0.1$, the mass of $Z_F$ is greater than 
3 TeV from current data~\cite{atlas14_z,cms15_z}.}
\item {The 125 GeV particle observed at the LHC is identified as $h$ which 
does not exactly correspond to the one Higgs boson of the SM.  In the 
simplest scenario studied in this paper, it should have only diagonal 
couplings to quarks and leptons as in the SM, but with an additional 
overall factor $x_i$ for the different families.  Hence $h \to \mu^- \mu^+$ 
as well as $h \to c \bar{c}$ are allowed to be several times those of the 
SM, which will be probed with more data in the future; whereas 
$h \to \tau \mu$ is forbidden, despite a hint~\cite{cms15_tau} from 
the CMS Collaboration that it may be nonzero.  This absence of 
flavor-changing neutral-current couplings corresponds to a symmetry limit 
where $U_{uR} = U_{dR} = U_{lR} = 1$.}
\end{itemize}

\noindent \underline{\it Acknowledgement}~:~
This work was supported in part by the U.~S.~Department of Energy Grant 
No. DE-SC0008541.

\baselineskip 18pt
\bibliographystyle{unsrt}

\begin{thebibliography}{99}
\bibitem{m01} E. Ma, Phys. Rev. Lett. {\bf 86}, 2502 (2001).
\bibitem{glr09} W. Grimus, L. Lavoura, and B. Radovcic, Phys. Lett. 
{\bf B674}, 117 (2009).
\bibitem{km16} C. Kownacki and E. Ma, arXiv:1604.01148 [hep-ph].
\bibitem{atlas14_z} G. Aad {\it et al.}, (ATLAS Collaboration), Phys. Rev. 
{\bf D90}, 052005 (2014).
\bibitem{cms15_z} S. Khachatryan {\it et al.}, (CMS Collaboration), JHEP 
{\bf 1504}, 025 (2015).
\bibitem{atlas12} G. Aad {\it et al.}, (ATLAS Collaboration), Phys. Lett. 
{\bf B716}, 1 (2012).
\bibitem{cms12} S. Chatrchyan {\it et al.}, (CMS Collaboration), Phys. Lett. 
{\bf B716}, 30 (2012).
\bibitem{cms15_tau} V. Khachatryan {\it et al.}, (CMS Collaboration), 
Phys. Lett. {\bf B749}, 337 (2015).
\bibitem{atlas14_mu} G. Aad {\it et al.}, (ATLAS Collaboration), Phys. Lett. 
{\bf B738}, 68 (2014).
\bibitem{cms15_mu} V. Khachatryan {\it et al.}, (CMS Collaboration), 
Phys. Lett. {\bf B744}, 184 (2015).
\bibitem{dgps14} C. Delaunay {\it et al.}, Phys. Rev. {\bf D89}, 033014 
(2014).


\end{thebibliography}

\end{document}